\input psfig
\documentstyle[12pt]{article}

\def\e{\mbox{e}}

\def\half{{1 \over 2}}
\def\a{\alpha}
\def\b{\beta}
\def\k{\kappa}
\def\t{\Theta}

\catcode`\@=11
\long\def\@makefntext#1{
\protect\noindent \hbox to 3.2pt {\hskip-.9pt
$^{{\ninerm\@thefnmark}}$\hfil}#1\hfill}		%CAN BE USED

\def\@makefnmark{\hbox to 0pt{$^{\@thefnmark}$\hss}}  %ORIGINAL

\def\ps@myheadings{\let\@mkboth\@gobbletwo
\def\@oddhead{\hbox{}
\rightmark\hfil\ninerm\thepage}
\def\@oddfoot{}\def\@evenhead{\ninerm\thepage\hfil
\leftmark\hbox{}}\def\@evenfoot{}
\def\sectionmark##1{}\def\subsectionmark##1{}}

\renewcommand{\thefootnote}{\fnsymbol{footnote}}

\newcounter{sectionc}\newcounter{subsectionc}\newcounter{subsubsectionc}
\renewcommand{\section}[1] {\vspace*{0.6cm}\addtocounter{sectionc}{1}
\setcounter{subsectionc}{0}\setcounter{subsubsectionc}{0}\noindent
	{\normalsize\bf\thesectionc. #1}\par\vspace*{0.4cm}}
\renewcommand{\subsection}[1] {\vspace*{0.6cm}\addtocounter{subsectionc}{1}
	\setcounter{subsubsectionc}{0}\noindent
	{\normalsize\it\thesectionc.\thesubsectionc. #1}\par\vspace*{0.4cm}}
\renewcommand{\subsubsection}[1]
{\vspace*{0.6cm}\addtocounter{subsubsectionc}{1}
	\noindent {\normalsize\rm\thesectionc.\thesubsectionc.\thesubsubsectionc.
	#1}\par\vspace*{0.4cm}}

\newcounter{appendixc}
\newcounter{subappendixc}[appendixc]
\newcounter{subsubappendixc}[subappendixc]

\renewcommand{\appendix}[1] {\vspace*{0.6cm}
        \refstepcounter{appendixc}
        \setcounter{figure}{0}
        \setcounter{table}{0}
        \setcounter{equation}{0}
        \renewcommand{\thefigure}{\Alph{appendixc}.\arabic{figure}}
        \renewcommand{\thetable}{\Alph{appendixc}.\arabic{table}}
        \renewcommand{\theappendixc}{\Alph{appendixc}}
        \renewcommand{\theequation}{\Alph{appendixc}.\arabic{equation}}
        \noindent{\bf Appendix \theappendixc #1}\par\vspace*{0.4cm}}

\def\abstracts#1{{

\centering{\begin{minipage}{12.2truecm}\footnotesize\baselineskip=12pt\noindent
	\centerline{\footnotesize ABSTRACT}\vspace*{0.3cm}
	\parindent=0pt #1
	\end{minipage}}\par}}

\renewenvironment{thebibliography}[1]
	{\begin{list}{\arabic{enumi}.}
	{\usecounter{enumi}\setlength{\parsep}{0pt}
\setlength{\leftmargin 1.25cm}{\rightmargin 0pt}
	 \setlength{\itemsep}{0pt} \settowidth
	{\labelwidth}{#1.}\sloppy}}{\end{list}}

\newcounter{itemlistc}
\newcounter{romanlistc}
\newcounter{alphlistc}
\newcounter{arabiclistc}

\newcommand{\fcaption}[1]{
        \refstepcounter{figure}
        \setbox\@tempboxa = \hbox{\footnotesize Fig.~\thefigure. #1}
        \ifdim \wd\@tempboxa > 6in
           {\begin{center}
        \parbox{6in}{\footnotesize\baselineskip=12pt Fig.~\thefigure. #1}
            \end{center}}
        \else
             {\begin{center}
             {\footnotesize Fig.~\thefigure. #1}
              \end{center}}
        \fi}

\newcommand{\tcaption}[1]{
        \refstepcounter{table}
        \setbox\@tempboxa = \hbox{\footnotesize Table~\thetable. #1}
        \ifdim \wd\@tempboxa > 6in
           {\begin{center}
        \parbox{6in}{\footnotesize\baselineskip=12pt Table~\thetable. #1}
            \end{center}}
        \else
             {\begin{center}
             {\footnotesize Table~\thetable. #1}
              \end{center}}
        \fi}

\def\@citex[#1]#2{\if@filesw\immediate\write\@auxout
	{\string\citation{#2}}\fi
\def\@citea{}\@cite{\@for\@citeb:=#2\do
	{\@citea\def\@citea{,}\@ifundefined
	{b@\@citeb}{{\bf ?}\@warning
	{Citation `\@citeb' on page \thepage \space undefined}}
	{\csname b@\@citeb\endcsname}}}{#1}}

\newif\if@cghi
\def\cite{\@cghitrue\@ifnextchar [{\@tempswatrue
	\@citex}{\@tempswafalse\@citex[]}}
\def\citelow{\@cghifalse\@ifnextchar [{\@tempswatrue
	\@citex}{\@tempswafalse\@citex[]}}
\def\@cite#1#2{{$\null^{#1}$\if@tempswa\typeout
	{IJCGA warning: optional citation argument
	ignored: `#2'} \fi}}

 1
 1
 1

\font\ninerm=cmr9

\begin{document}

\centerline{\normalsize\bf SEARCH FOR SOLITONS IN TWO-HIGGS }
\baselineskip=16pt
\centerline{\normalsize\bf EXTENSIONS OF THE STANDARD MODEL
\footnote{To
appear in the proceedings of the conference: "Topics in Quantum Field
Theory: Modern Methods in Fundamental Physics",
Maynooth, Ireland, May 1995. Ed. D. H. Tchrakian, World Scientific 1995.}
}
\baselineskip=15pt

%\vfill
\vspace*{0.6cm}
 \centerline{\footnotesize T. N. TOMARAS \footnote{I am grateful
to C. Bachas for a most fruitful and enjoyable collaboration throughout
this work.} }
\baselineskip=13pt
\centerline{\footnotesize\it Department of Physics, University of Crete, and
Research Center of Crete}
\baselineskip=12pt
\centerline{\footnotesize\it 714 09 Heraklion, Crete, Greece}
\centerline{\footnotesize E-mail: tomaras@plato.iesl.forth.gr}
%\vspace*{0.3cm}
\vspace*{0.9cm}
\abstracts{ We report on the status of
our search for quasi-topological
solitons of various dimensions in realistic field theoretical models of
condensed matter and
of elementary particle physics.
}

\section{Introduction}
\indent
%\vfil
%\vskip 0.8cm
%\rm\baselineskip=14pt
The search for stable lumps in the Weinberg-Salam model
has a long history.
It has revealed a rich
structure of classical solutions including the
sphaleron\cite{DHN,Taubes,Klink,Boguta} ,
deformed sphalerons\cite{eila,Brihaye,Yaffe,Klink2}
 and vortex strings\cite{Nambu,Soni,Vach,Vach1,Periv}.
Such solutions could play a role in understanding
{\ninerm (B+L)}-violation
and structure formation in the early universe,
but they are all classically-unstable or/and extended.
They have therefore
no direct  present-day manifestation,
contrary to long-lived particles whose
relic density could at least in principle be
detected.

The existence of particle-like excitations has, on the
other hand, been argued for in the context of a strongly-
interacting Higgs sector\cite{Tze,Fahri,Rub,Zee,Carlson}.
The advocated particles can be thought of as
technibaryons of
an underlying technicolor model.
They are described in
bosonic language by winding solitons of
 an effective non-renormalizable
lagrangian for the   pseudo-goldstone-boson
 (or technipion) field,
much like  skyrmions\cite{Skyrme}
 of the effective chiral lagrangian
of $QCD$.  This is of course a phenomenological
 description, since
the   properties of such hypothetical
particles cannot be
calculated reliably within a semi-classical expansion.
Furthermore, in
view of the difficulties
facing technicolor models,
the  possibility of a strongly-interacting Higgs sector is
not theoretically  appealing.

It would be clearly    more interesting
 if classically-stable winding excitations
could  arise in a {\it weakly-coupled} scalar
 sector.
To be more precise let us decompose the
 Higgs-doublet  field into a
real (positive)  magnitude   and a
group-phase: $\Phi = F U$, and
consider static configurations with $U(x)$
 wrapping $N$ times
around the ${\rm SU(2)}$ manifold. These are
 potentially unstable
for at least three distinct reasons:
%\break
{\it (a)} because $N$ is not   conserved
whenever the magnitude $F$   goes through
zero;
%\break
{\it (b)} because $N$ is not gauge-invariant and can,
in particular, be non-vanishing
even in a vacuum state;
and {\it (c)} because scalar-field configurations can loose
their energy
by shrinking to zero size\cite{Derrick}.
We refer to these for short as the {\it radial, gauge}
 and {\it scale}
instabilities. They can be eliminated formally by
{\it (a)}
 taking the physical-Higgs mass $m_H\to\infty$,
{\it (b)} decoupling the electroweak gauge fields,
 and
{\it (c)} adding appropriate higher-derivative terms to the
 action.
The question is whether classical stability can
 be maintained
while relaxing the above conditions.
This
 has been investigated numerically in the
{\it minimal} case of one
doublet: although one may indeed relax both
the weak gauge coupling\cite{Rub,eila}
and  the Higgs mass\cite{Kunz}
  up to some finite critical values,
  stability cannot apparently be achieved without the
   non-renormalizable  higher-derivative
  terms in the
 action.
On the other hand, as we have
 demonstrated recently,
  metastable winding solitons do arise in
renormalizable models in
two\cite{mexican} and three\cite{preprint}
space-time dimensions. The way this happens
is the content of section 2.
It is we believe instructive and could guide
the search for lower dimensional structures in
many physical systems as well as for similar
semi-classical solitons in four dimensions.
Section 3 contains a brief
presentation of the status of our search for particle-like solutions
in the context of two-Higgs extensions of the standard model.
We close with some comments in the discussion section.

\setcounter{footnote}{0}
\renewcommand{\thefootnote}{\alph{footnote}}

\section{Quasi-topological solitons in lower dimensions }

The simplest context in which the {\it radial}
instability is an issue is a two-dimensional model of
a complex scalar field with mexican-hat potential:
$ V = {1\over 4}\lambda (\Phi^*\Phi - v^2)^2$.
The simplest way to describe in detail the winding solitons
is to take space to be periodic with period
$L$ \footnote{Alternatively we may add a mass term:
$\delta V = -\mu^2 v Re(\Phi)$,  that
lifts the vacuum degeneracy. Stable winding excitations, which reduce to
the sine-Gordon solitons in the $\lambda\to\infty$ limit,
can be shown\cite{Pallis} numerically to exist for
$\lambda v^2/\mu > 18.8$ \  .} .
 The condition for classical stability can in this case
be derived analytically and reads\cite{mexican} : $m_H L > \sqrt{5} \ ,$
where $m_H=\sqrt{2\lambda} v$. The classically-relevant parameter
is thus the radial-Higgs mass in units of the soliton  size.
This follows also by comparing  the loss in potential energy to the gain
in gradient energy when trying to undo the winding by
reducing  the magnitude of the scalar. Note that
the loop-expansion parameter $\lambda L^2$, can be taken to zero
independently so as to reach a semiclassical limit.

 The above winding solitons become unstable
  classically if we gauge the $U(1)$ symmetry of the model.
The {\it gauge} instability is in fact
more severe than in four dimensions,
because   no energetic barrier
  opposes the turning-on of
a static space-like gauge field, which is necessary
 to reach a winding-vacuum
state. The minimal Abelian-Higgs model has thus only unstable
(sphaleron) solutions\cite{Giller}
 \footnote{It was claimed erroneously in [22] that it
has no static solutions whatsoever.
 This is only correct in the $\lambda\to\infty$
limit.}$\ $ .
The situation changes, however, drastically
if there are more than one complex scalars. The
gauge-invariant  relative phases
of any two of them   cannot in this case  wind around
non-trivially in a vacuum state.
 An explicit
analysis
of this extended abelian-Higgs model\cite{mexican}
 shows that   winding solitons
  persist down to scalar masses close to
 the inverse soliton size: gauging and the extra Higgs
  enhance the stability region found in the global model.

 The {\it scale} instability becomes an issue for
 the first time
in three space-time dimensions.
To be more precise we consider a real-triplet
 scalar field $\Phi_a(x)$
($a=1,2,3$) with mexican-hat
potential : $V= {1\over 4}\lambda (\Phi_a\Phi_a - v^2)^2$.
The  limit $\lambda\to\infty$ corresponds to the
$O(3)$
non-linear $\sigma$-model. This is known to possess
 winding solitons, characterized by non-trivial
 mappings of the two-sphere
onto itself, and having
  arbitrary size\cite{Belavin}.
For finite $\lambda$ on the other hand, or in the
 presence of a
symmetry-breaking potential, Derrick's
 scaling argument\cite{Derrick}
 shows that these   solitons are unstable to shrinking.
One can of course again invoke higher-derivative terms
to stabilize the scale\cite{Zak}.
The same result is however in this case  achieved by a massive
$U(1)$ gauge field with only renormalizable
 couplings\cite{preprint}.
This can be established
 by perturbing around the
$O(3)$ non-linear $\sigma$-model limit,
 or else by solving numerically
the equations of motion.

What do these lower-dimensional solitons teach us?
{\it First}, they suggest by analogy that
classically-stable winding solitons may
exist in a weakly-coupled two-Higgs extension of the standard model.
The status of our search for such localized solutions is the content
of the following section.

{\it Second}, they are interesting in their own right,
since they correspond
to a new class of wall and string defects in realistic
four-dimensional models of particle and condensed-matter
physics . One such example are the metastable
{\it membranes}, discovered recently\cite{membranes,Dvali} in
two or more Higgs-doublet extensions of the standard model.
These are static
wall-type solutions, non-topological but classically stable
in a wide region of the Higgs sector
parameter space compatible with
perturbative unitarity and with present phenomenological bounds.
They are embeddings of the above discussed\cite{mexican}
solutions of the 2d Abelian-Higgs model with two or more complex scalars.
They are characterized by the non-trivial winding of
the relative U(1) phase of the neutral components of any
two Higgs doublets of the extended standard model in the
direction $x$ normal to the wall. They have no electromagnetic coupling,
their size is a few times the inverse of the mass $m_A$
of the CP-odd scalar $A^0$ (in the standard notation of the generic
two-Higgs model\cite{vivlio}), while their energy per unit area
is in terms of the $A^0$ and W masses and the fine structure constant
of order $m_W^2 m_A /\alpha$.
Assuming a $m_A \simeq 50 GeV$ i.e. not much larger than
the present experimental lower bound\cite{vivlio}, the mass
of the wall is of ${\cal O} (10^{10} gr/cm^2)$. Thus, a single
wall crossing the entire Universe today would by far overclose it
and can be excluded. Smaller membranes though, which either
collapsed or were torn apart by quantum tunneling may have acted
as seeds for the formation of galaxies. In fact, the mass of
a typical galaxy is comparable to that of a membrane a few
light years in size. In any case, the production and decay
rates of these objects and their cosmological role
has to be studied in detail, since most likely
they are going to lead to firm constaints on the Higgs mass
spectrum of extensions of the standard model\cite{membranes}.

As another example, we would like to mention the static
wall-type metastable solutions of the easy-plane
ferromagnetic continuum in the presence of an external
in-plane magnetic field $h$ \footnote{The search for finite-energy
soliton solutions
in ferromagnetic and antiferromagnetic systems and the study of their
stability, reported briefly in
this paragraph,
was carried out in collaboration with Prof. N. Papanicolaou.
I would also like to thank Dr. P. Tinyakov for helpful discussions
about sphaleron
solutions in relativistic anisotropic
non-linear $\sigma-$models in 1+1-dimensions.} .
The dynamics of the system is described by the Landau-Lifchitz
equations with energy density
given in terms of the unit magnetization vector $n_a, \;a=1,2,3,$
by:
$
{\cal E} = {1\over2} (\partial_i n_a)^2 +
g {n_3}^2 + h (1 - n_1).
$
The anisotropy constant $g$ and the external field $h$ are
both taken to be positive.
The system has the unique semiclassical ground state:
$(n_1=1, n_2=0=n_3)$ and
no topologically stable domain walls.
Nevertheless, it posesses a variety of wall-type static finite
energy (per unit area) solutions. A particularly interesting one is
$(n_1=cos\Theta(x), n_2=sin\Theta(x), n_3=0)$,
with $\Theta(x)$ satisfying $\Theta^{\prime \prime} (x) = h sin\Theta(x)$
and the boundary conditions $\Theta(-\infty)=0, \Theta(+\infty)=2 \pi$.
It can be verified that this solution is
dynamically stable for $g/h \ge 3$.

\section{Localized solutions in 3+1 dimensions
\footnote{The work in this section was
done in collaboration with Dr. P. Tinyakov and
is described in detail in reference [31].} }

We now turn our attention to the possibility of classically stable
localized particle-like winding
solitons in 3+1 dimensions.
In the context of a two-Higgs extension of the standard model
these hypothetical solitons would: {\it (b)} be characterized
by the non-trivial winding
of the relative phase of the two doublets, and thus be immune to
the gauge mode of decay; {\it (c)} have a scale stabilized by
electroweak magnetic fields and hence  of order $1/m_W$;
and {\it (a)}   hopefully stay stable for Higgs masses near $m_W$
and thus compatible with perturbative unitarity
 \footnote{Though admittedly premature, some other
physical properties of such would-be particles are fun to
contemplate: being classically stable they could easily have
cosmological life times. They would have a mass in the $\sim 10\  TeV$
region, zero charge and dipole moments in their ground state,
and geometrical interaction cross sections of order $1/m_W^2$.
Assuming maximum production at the electroweak phase transition,
a rough estimate of their present abundance shows that they
could be candidates for cold dark matter in the
universe.}$\ $.
Mathematically the situation is the same as in the
hidden-gauge-boson models\cite{Kunzz,Dobado} of strong
 and electroweak interactions,
except that the role of the hidden gauge bosons is here played
by  $W^{\pm}$ and $Z$ themselves.

%%%%%%%%%%%%%%%%%%%%%%%%%%%%%%%%%%%%%%%%%%%%%%%%%%%%%%%%%%%%%%%%%%%%%
%%%%%%%%%%%%%%%%%%%%%%%%%%%%%%%%%%%%%%%%%%%%%%%%%%%%%%%%%%%%%%%%%%%%

We carried out a numerical search\cite{Kunzz,Peccei,Tinyakov}
for such stable spherically symmetric particle-like
soliton solutions in a simplified version of
the two-Higgs $\rm SU(2)\times U(1)$ model.
The presumably inessential simplification consists
of taking (a) the $\rm U(1)$ gauge
coupling $g^\prime=0$ as required
by spherical symmetry and (b) the Higgs potential to be the sum
of two "mexican hats" one for each Higgs doublet.
The resulting action is:
\begin{equation}
  S={1\over {g^2}}\int d^4x \Bigl( -\half Tr(W_{\mu\nu}W^{\mu\nu})
  +\sum_{I=1,2} (D_{\mu} H_I)^{\dagger}(D^{\mu} H_I)
  +\sum_{I=1,2} {{\lambda_I}\over {g^2}}
  (H^{\dagger}_I H_I- g^2 v^2_I)^2 \Bigr)
\end{equation}
with $ D_\mu H_I=(\partial_\mu + W_\mu) H_I ,\;
I=1,2 $
the covariant derivative on
the Higgs doublets. $W_\mu \equiv {1\over{2i}} \tau^a W^a_\mu$ and
$W_{\mu\nu}$ is the ${\rm SU(2)}$ field strength.
We used the spherically symmetric
ansatz
  \[
  W_0=a_0\tau_in_i/2i\; ,
  \]
  \[
  W_i=[ (\alpha-1)\epsilon_{ijk}\tau_jn_k/r +
      \beta(\delta_{ij}-n_in_j)\tau_j/r + a_1n_in_j\tau_j ]/2i
  \]
  \[
  H_I=(\mu_I+i\nu_I n_i\tau_i)\xi
  \]
where $n_i$ is the unit vector in the direction of $\bf x$, $\tau_i$ are the
three Pauli matrices, $\xi$ is a constant unit doublet and $a_0, a_1,
\alpha, \beta, \mu_I, \nu_I$ are functions of $r$ to be determined
by the energy extremization. It is convenient to choose the
gauge $a_0=0$. One then solves Gauss' constraint
to obtain $a_1=0$\cite{Yaffe}.

We rescale distances and fields by the appropriate
powers of $m_W=g \sqrt{(v_1^2+v_2^2)/2}$ and use the notation
$\phi_I=\mu_I+i\nu_I \equiv F_I \e^{i\Theta_I}$ to write for the
energy functional:
  \[
  E = {M_W \over \alpha_W}
  {2\over 1+t^2} \int dr \Bigl\{
  {1+t^2\over 2}(\a'^2+\b'^2)
  + r^2(F_1'^2+F_2'^2+F_1^2\t_1'^2+F_2^2\t_2'^2)
  \] \nopagebreak
  \[
  + {1+t^2\over 4r^2}(\a^2+\b^2-1)^2
  +\half (\a^2+\b^2+1)(F_1^2+F_2^2)
  \] \nopagebreak
  \[
  -\a[F_1^2\cos(2\t_1)+F_2^2\cos(2\t_2)]
  -\b[F_1^2\sin(2\t_1)+F_2^2\sin(2\t_2)]
  \] \nopagebreak
  \begin{equation}
  +{\k_1^2\over 4}r^2(F_1^2-1)^2
  +{\k_2^2\over 4t^2}r^2(F_2^2-t^2)^2 \Bigr\},
  \label{E'}
  \end{equation}
where $t\equiv {v_2}/{v_1}$ and $\k_I\equiv {m_{H_I}}/m_W$. The
radial coordinate in particular in the above expression
is measured in units of $m_W^{-1}$.

We require finiteness of the energy of the solution and use
the global symmetries of the energy functional\footnote{The energy
functional is invariant under the simultaneous rotation of the
complex fields $\phi_1, \phi_2 \;{\rm and}\; \alpha+i \beta$ by an angle
$\omega, \omega \;{\rm and}\; 2 \omega$ respectively, as well as
under independent rotations of $\phi_I$ by $\pi$.}$\;$ to specify the
boundary conditions at $\infty$ and write
them in the form:
  \[
  \a\to 1, \;\;\;\;
%  \]
%  \[
  \b\to 0, \;\;\;\;
%  \]
%  \[
  F_1\to 1, \;\;\;\;
%  \]
%  \[
  F_2\to t, \;\;\;\;
%  \]
%  \[
  \t_a\to 0
  \]
Correspondingly, at $r=0$ energy finiteness implies
  \[
  \a^2+\b^2\to 1.
  \]
while the field equations obtained by varrying (3)
with respect to $\Theta_I$, together with the requirement
of smoothness of the solution lead to the dynamical
conditions:
$
2 \Theta_a - \psi = 0 \;{\rm mod} \;2 \pi
$
, where $\psi$ is the phase of the
field $\chi \equiv \alpha+i \beta$.
Thus all solutions satisfy $  \Theta_1 -  \Theta_2 =  \pi N$
at the origin.

We have looked for solutions in a wide region of the
$\{t, \kappa_1, \kappa_2 \}$ parameter space.
We refer the reader to [31] for
the detailed presentation of the numerical method we used and of
the entire solution-zoo we have obtained so far.
In the present note, I will
describe briefly some of the solutions we found
for various values of $\kappa$ on
the line $\{ t=1, \kappa_1=\kappa_2 \equiv \kappa \}$,
together
with their main characteristics.
As we varry $\kappa$ the energy landscape changes smoothly
and consequently, one generically
expects a continuous change in the number and
properties of its extrema and
saddles.
Figure 1 below captures these changes in part of the energy
landscape and shows some of the solutions of the model at hand.

%\hbox{
%\hspace{0.57cm}
%\psfig{file=graphtree.ps,bbllx=50bp,bblly=90bp,bburx=500bp,bbury=790bp,height=8cm,width=15cm,angle=270}
%}

%\vspace {1.5cm}

The graph should be thought of as an "artist's conception" of a part
of the tree
of solutions with $N=0$ and $|N|=1$ which has emmerged so far
in the two-Higgs
model under study. Each branch of the "tree" corresponds to
a particular type of solution. As we move along a branch by varrying
$\kappa$, continuous quantities like the detailed
form of the solution, its energy as well as the value(s) of the
imaginary eigen-frequencies of the small oscillations around
it change smoothly, while integer-valued ones
such as the index $N$, the number of unstable negative curvature modes
(shown on the branch) and
the number of nodes of the fields, remain invariant.
We have suppressed the branches corresponding to the
CP-conjugates of the solutions shown. They may be
imagined as the mirror images with respect to the
$\kappa-$axis of the branches above it.

For $\kappa$ between 0.0 and 5.5 the only non-trivial solution is
the sphaleron embedded in the two-Higgs model;
it has $\Theta_1=\Theta_2=0$, $\beta=0$
and one negative mode. For $\kappa$ about 5.5 we reach the
bifurcation point ${\rm B}_0$, the sphaleron
develops a second negative mode and at the same time two new solutions,
namely the
branch $A_1$ and its CP conjugate
appear with $N=1$ and $N=-1$ respectively.
The phases $\Theta_1$ and $\Theta_2$ and the complex field
$\chi$ of the solution corresponding to the branch $A_1$
have the behaviour we call
of "type-A" and show in figure 2.

%\vspace {1.cm}
%\hbox{
%\hspace {0.7cm}
%\psfig{file=graphTypeAS.ps,bbllx=50bp,bblly=260bp,bburx=500bp,bbury=500bp,height=6.cm,width=11.5cm,angle=0}

%}

We have checked that the branch ${\rm A}_1$ extends without
a new bifurcation
up to $\kappa=200$. As for the sphaleron, once we get to
$\kappa \simeq 12.0$ we encounter the bifurcation ${\rm B}_1$.
The sphaleron aquires a third unstable mode
and two new solutions (branch ${\rm S}_2$ and its CP-conjugate) emerge.
They have $N=0$ and phase behaviour of "type-S" shown above.
And so on for the rest of the tree.

Incidentally, using the number of negative modes $N_{\rm neg. \;modes}$
of the solutions
shown in figure 1 one may explicitly verify
the sum rule:
$
\sum_{\rm solutions} (-)^{N_{\rm neg. \;modes}} =
{\rm constant}
$
independent of $\kappa$,
for the whole range of values we considered.

\section{Discussion}

It is clear that the two-Higgs system discussed here posesses
a variety of unstable solutions,
whose richness
increases with the Higgs masses.
These, like the ordinary sphaleron, are in principle interesting in their
own right especially in connection with the baryon number generation
at the electroweak phase transition.
There are important differences\cite{Tinyakov}
in the values of the energies
and of the negative modes of the above solutions
compared to analogous ones
of the minimal standard
model.
For instance, the value of the negative
curvature along the $A_1$
branch is about half of what it is
in the corresponding least unstable deformed
sphaleron of the one-Higgs model. These differences
affect directly the
prediction for the baryon number in the Universe, and at the same time
they offer support to our general arguments
that an extended Higgs sector improves
the stability of the winding solitons under discussion.

Despite of this indirect evidence though, we have not as yet
been able to find a stable soliton.
Such a solution might for example arise at
a bifurcation point like X of figure 1
and either merge with another branch of
the tree at some new bifurcation
or continue, as is the case with $A_1$, up to $\kappa=\infty$.
But since our numerical method requires an initial
guess for the solution which has to lie inside the basin of attraction
of the corresponding
local minimum of the energy functional,
our failure to find a stable solution might just be
due to the bad starting configurations we tried so far. Further
theoretical analysis is required to guide our numerical search.

%%%%%%%%%%%%%%%%%%%%%%%%%%%%%%%%%%%%%%%%%%%%%%%%%%%%%%%%%%%%%%%%%
\vspace{1cm}
{\bf Acknowledgements}
\vskip 0.4cm

This research was supported in part by
the EEC grants CHRX-CT94-0621 and CHRX-CT93-0340,
as well as by the Greek General Secretariat
of Research and Technology
grant 91$E\Delta$358.

%\endpage

\end{document}